\documentclass[12pt,twoside,english,aps]{revtex4}
\usepackage[T1]{fontenc}
\usepackage[latin9]{inputenc}
\usepackage{amsmath}
\usepackage{amssymb}
\usepackage{graphicx}

\makeatletter
\@ifundefined{textcolor}{}
{%
 \definecolor{BLACK}{gray}{0}
 \definecolor{WHITE}{gray}{1}
 \definecolor{RED}{rgb}{1,0,0}
 \definecolor{GREEN}{rgb}{0,1,0}
 \definecolor{BLUE}{rgb}{0,0,1}
 \definecolor{CYAN}{cmyk}{1,0,0,0}
 \definecolor{MAGENTA}{cmyk}{0,1,0,0}
 \definecolor{YELLOW}{cmyk}{0,0,1,0}
}

\usepackage{epsfig}\usepackage{color}\usepackage{multirow}

\usepackage[english]{babel}

\makeatother

\usepackage{babel}
\begin{document}

\title{ Ultrametric dynamics for the closed fractal-cluster resource models}

\author{V.T. Volov}

\affiliation{Physics Department, Samara State University of Railway Transport,
Samara, Russia, e-mail: vtvolov@mail.ru}

\author{A.P. Zubarev}

\affiliation{Physics Department, Samara State University of Railway Transport,
Samara, Russia, e-mail: apzubarev@mail.ru}

\date{\today}
\begin{abstract}
The evolutional scenario of the resource distribution in the fractal-cluster
systems which is identified as an "organism" has been suggested.
We propose a model in which the resource redistribution dynamics in
the closed system is determined by the ultrametric structure of the
system's space. Moreover, each cluster has its own character time
of a transfer to the equilibrium state which is determined by the
ultrametric size of the cluster. The general equation which determines
this dynamics has been written. For the determined type of the resource
transitions among clusters, the solution to this equation has been
numerically received. The problem of the parameter's identification
modeling for the real systems has been discussed.
\end{abstract}
\maketitle
%

\section{Introduction}

\label{sect:1}

The progress of mathematics for modeling systems which have a visible
or hidden hierarchical structure is important for the wide class of
systems and processes in different spheres of physics, biology, economics
and sociology: the spin glasses, biopolymers, the theory of optimization,
the taxonomy, the evolutional biology, the cluster and factor analysis
and etc. \cite{RTV,ALL}.

In fact any biological or socio-economic system has an evident hierarchical
character of interaction among its subsystems and, this way, carries
inside itself elements of the hierarchical structure. If the initial
objects or states of the system have the hierarchical structure then
the structure is visible. However, all systems with a hidden hierarchical
structure present more interest for the researchers. In these systems
the hierarchical structure is not sought in the initial variables,
but they have been observed after transition to some effective variables.
As a rule, the number of these effective variables is essentially smaller than the number
of the degrees of freedom for the whole system. Well-known members
of the same systems are spin glasses \cite{Dot}, proteins \cite{OS,ABKO,ABO}.
There are reasons for the identification of hierarchical structures
among the social-economical systems \cite{SJ,MS,BZK}. Adequate mathematical
apparatus for the modelling systems which can be both visible and
hidden hierarchical structures is the ultrametric analysis \cite{ALL,S,VVZ,Sh_Kh,DZ}.

One of the interesting facts which we have observed on the empirical
researches, which relates to the reasonably wide class of complex
natural and artificial genetic systems, is the certain system's type
being manifested. These systems have a strongly marked hierarchical
structure for the recourse distribution on the functional indication
\cite{Burd}. It has been observed in a range of the systems (biological,
technical, social) which have been evolutied for billions of years,
but are now in a stable functioning state. These systems have five
basic subsystems (cluster) which can be classified with their target
functions: energy, transport, technological, ecological and informational
clusters, where each of the ones have a certain share of "the resource".
For such systems, the special term -- "an organism"  has been
determined. For the social-economical systems which belong to the
special class of "an organism" and it is these systems which have
the most interest for the investigation in this article. The clusters
will be identified on the target distribution for the extensive parameters
as staff, industry funds, financial activitias etc.

For the range of systems which relate to the class of "the organisms",
each cluster can present itself so the functioning subsystem which
is "an organism" too. Ergo we can divide the one to five subsystems
(subclusters), which have the same target functions for each system
inside itself of the one. For example, the resource in the energetic
cluster can be to share for the own energetic, transport, ecology,
technological and informational supporting. Such a decomposition can
be extended to subclusters. Namely, each of the subclusters of given
level can be considered as the union of the five high level subclusters.
Such systems are called fractal-cluster systems (see \cite{Volov1,Volov2,Volov3}).
Thus the space of the resource distribution for the same systems has
an hierarchical structure which can be discribed by the hierarchical
tree with a fixed number of branches $p$=5. It should be noted that
identification of the subcluster in the particular fractal-cluster
system dependens on the type of "the resource", which has not
always had a clear relationship to the observed characteristics of
the system. Practically under the fractal-cluster modeling of the
real social-economical, biological, technological and other systems
of this type we have the possibility to describe only two or three
hierarchical levels of the ones of the clusterization. Even such a
classification of the resource distribution allows the production
of the functioning estimation effectiveness for any real complex systems.
Nevertheless, by using the mathematical apparatus, we can research
the resource dynamics in the abstract systems with an infinite number
of nested clusters (subclusters).

Statistical analysis of the empirical data on the resource distribution
in the complex social-economical, technological, biological and other
systems of the ones type allows to determine the ideal values for
the resource distribution in the first level clusters \cite{Volov1,Volov2,Volov3}.
Under these values, the system's development will be stable and more
energy efficient. First of all, in the anthropological and technological
systems, the statistical analysis allows to obtain the average values
of the ideal distribution. These values for energetical, transport,
ecological, technological and informational clusters are equal $0.38,\:0.27,\:0.16,\:0.13,\:0.06$
respectively. For the system of another nature (biological, social
and economical) these ideal volumes have approximatly the same values.

The possibility of using the fractal-cluster models for the resource
distribution analysis is based on the researches of the range investigations
\cite{Volov1,Volov2,Volov3}, in which researches the optimal resource distribution
control methods had been presented. These methods are based on the
formal analogies with the thermodynamic method in its informational
interpretation. Such an approach has allowed us to receive the solutions
close to the ideal resource distribution state. However, we have an
open question -- what is the way the same systems which are placed
out the equilibrium (the ideal state) go to the ideal state soon.
In this article we suggest the scenario of the closed fractal-cluster
system's evolution from the arbitrary state to the ideal state. We
have postulated that dynamics of the resource redistribution in the
closed systems under the condition of the external control factors
absence is wholly determined by the ultrametric structure of the fractal-cluster
space inside of the one this resource is allocated. It has been shown
that for each cluster we have own characteristic time transition to
the ideal substate. The characteristic time transition is determined
by the ultrametric size of the cluster (subcluster). From the beginning,
the subclusters of the highest levels transfer to the ideal state,
then the ones of more lower level make the same action and so on.
This hierarchical structure of the characteristic transition time
of all subclusters to the ideal substates has been determined for
the whole system's dynamics. The general equation which describes
such dynamics has been written. For certain types of resource transitions
among the clusters we have numerical researches for solving this equation.
Also we have discussed the problem of the parameter model's identification
for the real systems.

\section{Model of resource distribution dynamics for the closed fractal-cluster
systems}

\label{sect:2}

We are considering a graph which is the $n$-level hierarchical tree
with the one root vertex, which is the center of the graph. Let $p$
be the index of the tree branching, $n$ is the hierarchical level
number.

\begin{figure}[ht]
\begin{centering}
\includegraphics[clip,width=0.7\linewidth]{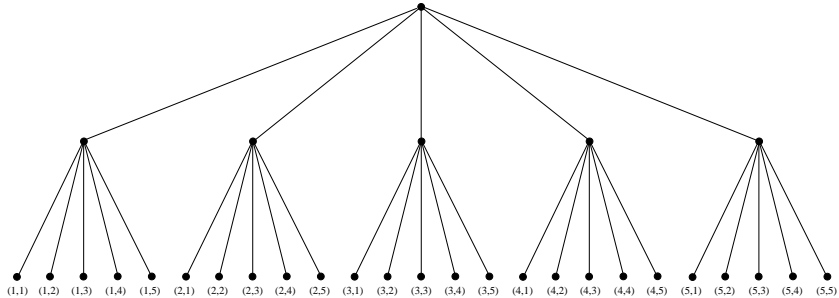} \caption{Hierarchical tree which corresponds to the two level the fractal-cluster
system. Here $(a_{1},a_{2})$, $a_{1},\: a_{2}=1,\ldots,p$, $p=5$
is the boundary points (the second level clusters) tree parameterization
. }

\par\end{centering}

\label{tree}
\end{figure}

The sample of this tree has presented on the fig. \ref{tree} (here
$p=5$, $n=2$). The ensemble of the boundary points tree has indentificated
as $U_{n}$. Let $x$ is a point of the boundary. Then assignment
$x$ is equivalent to setting
\[
x=\left(a_{1},a_{2},\ldots,a_{n}\right),
\]
where $a_{i}=1,\ldots,p$, $i=1,\ldots,n$.

On the $U_{n}$ the ultrametric distance $d$ is putting. Namely,
for any two points $x=\left(a_{1},a_{2},\ldots,a_{n}\right)$ and
$y=\left(b_{1},b_{2},\ldots,b_{n}\right)$ one has:
\[
d(x,y)=d\left(a_{1},a_{2},\ldots,a_{n}|b_{1},b_{2},\ldots,b_{n}\right)=p^{n-\gamma+1}.
\]
Here $\gamma$ is determined by comparing the sets $\left(a_{1},a_{2},\ldots,a_{n}\right)$
and $\left(b_{1},b_{2},\ldots,b_{n}\right)$: if $a_{1}=b_{1}$, $a_{2}=a_{2},$$\ldots$,
$a_{j-1}=b_{j-1},$$a_{j}\neq b_{j}$, then it`s took $\gamma=j$.
For the similar elements this distance equals to zero.

As it is known \cite{S,VVZ}, any element $x$ of the $p$- adic numbers
field $Q_{p}$ can be presented in the following way $x=p^{-\gamma}\left(b_{0}+b_{1}p+b_{2}p^{2}+\cdots+b_{n}p^{n}+\cdots\right)$,
where $\gamma\in\mathbb{Z}$, $b_{0}=1,\ldots,p-1,$ and for $i\neq0$
$b_{i}=0,\ldots,p-1$. $p$-adic norm of the element $x$ is given
as $|x|_{p}=p^{\gamma}.$ For two elements $x,y\in Q_{p}$ distance
between them is the ultrametric and is determined as $d(x,y)=|x-y|_{p}.$
The subset $B_{n}=\left\{ x\in\mathit{Q_{p}}:\:|x|_{p}\leq p^{n}\right\} $
named the ball of radius $p^{n}$ in $p$-adic numbers $\mathcal{\mathit{Q_{p}}}$.
The subset $B_{n}/B_{0}$ consists from the elements of view $p^{-n}\left(b_{0}+b_{1}p+b_{2}p^{2}+\cdots+b_{n-1}p^{n-1}\right)$.
It has one-to-one correspondence between the subset $B_{n}/B_{0}$
of the $p$-adic numbers field $Q_{p}$ and the set of boundary points
of the tree $U_{n}$:
\[
U_{n}\ni\left(a_{1},a_{2},...,a_{n}\right)\longleftrightarrow p^{-n}\left(b_{0}+b_{1}p+b_{3}p^{2}+\cdots+b_{n-1}p^{n-1}\right)\in B_{n}/B_{0}\text{,}
\]
\[
b_{0}=a_{1}-1,\: b_{1}=a_{2}-1,\:\cdots\: b_{n-1}=a_{n}.
\]
The $i$-level cluster, $i=1,\ldots,n$, will be to name the subsets
of the set $U_{i}$, $i=0,1,\ldots,n-1$, such that $\forall\: x,\: y\in U_{n}$
$d(x,y)\leq p^{i}$. Any point $x\equiv\left(a_{1},a_{2},\ldots,a_{n}\right)\in U_{n}$
is the cluster $U_{0}=U_{0}\left(a_{1},a_{2},\ldots,a_{n}\right)$,
such clusters (top points of the ultrametric tree) we will be to name
by the highest level clusters or simply by points.

Let $F$ is some extensive parameter of the fractal-cluster system.
The distribution function $f(x)$ of the on the fractal-cluster space
$U_{n}$ will be to name the non-negative function $f(x)\equiv f(a_{1},a_{2},\ldots,a_{n})$,
which satisfies the following condition:
\[
\sum_{x\in U_{n}}f(x)=1,
\]
therefore the value $F_{i}=\sum_{x\in U_{i}}f(x)$ is the value $F$,
which relates to the cluster $U_{i}$.

Let one has non-negative numbers $q_{1},q_{2},\ldots,q_{p}$, which
satisfies to the following condition:
\[
\sum_{a=1}^{p}q_{a}=1.
\]
We will be to name the distribution $f^{ss}(a_{1},a_{2},\ldots,a_{n})$
the self-similar if

\begin{equation}
f^{ss}(a_{1},a_{2},\ldots,a_{n})=q_{a_{1}}q_{a_{2}}\ldots q_{a_{n}}.\label{ss}
\end{equation}
For the case $p=5$ we will be to name a self-similar distribution
$f^{id}(a_{1},a_{2},\ldots,a_{n})=q_{a_{1}}^{id}q_{a_{2}}^{id}\ldots q_{a_{n}}^{id}$
by the ideal one if
\[
q_{1}^{id}=0.38,\: q_{2}^{id}=0.27,\: q_{3}^{id}=0.16,\: q_{4}^{id}=0.13,\: q_{5}^{id}=0.06.
\]

Next we will research the model of the fractal-cluster system's evolution
which is estimated by the distribution $f(x,t)=f(a_{1},a_{2},\ldots,a_{n},t)$,
which depends from the time $t$. We will assume the following suggestions:

1. For any moment of the time $t$ the whole system's resource is
constant:
\begin{equation}
\sum_{x\in U_{n}}f(x,t)=1.\label{closed}
\end{equation}
If the suggestion (\ref{closed}) is realized for the some time interval
then this system is named as the closed system relatively of this
resource for this time interval.

2. For any initial distribution the system transfers to the ideal
state $f^{id}(x)$:
\begin{equation}
\lim_{t\rightarrow\infty}f(x,t)=f^{id}(x)\text{.}\label{st}
\end{equation}

3. The resource quantity which is transiting per the unit of time
from the highest level cluster (point) $y$ to any highest level cluster
(point) $x$ of the system decreases with increasing the ultrametric
distance $d(x,y)$ between these points:
\[
\dfrac{df(x,t)}{dt}\left|_{y\rightarrow x}\right.\sim K\left(d(x,y)\right)f(y),
\]
where $K(\lambda)$ is some positive decreasing function of the argument
$\lambda$.

These three suggestions allow to write of the dynamic equation for
the $f(x,t)$:
\begin{equation}
\frac{\partial}{\partial t}f(x,t)=\sum_{y\in U_{n},\: y\neq x}K\left(d(x,y)\right)\left(f^{id}(x)f(y,t)-f^{id}(y)f(x,t)\right).\label{Dyn}
\end{equation}
Summation of the right and left parts of the equation (\ref{Dyn})
on $x\in U_{n}$ gives us the conservation law of the whole system's
resource:
\[
\dfrac{\partial}{\partial t}\sum_{x\in U_{n}}f(x,t)=0.
\]
 Obviously the function $f^{id}(x)$ is the stationary solution (\ref{Dyn}).
This ensures that the condition (\ref{st}) is hold for any solution
$f(x,t).$

From the equation (\ref{Dyn}) it follows that the value
\begin{equation}
\tau_{i}=\dfrac{1}{K(p^{i})}\label{tau}
\end{equation}
is the characteristic transition time to the ideal state of the $i$-level
cluster i.e. the function
\begin{eqnarray*}
f(a_{1},\ldots,a_{i},t) & \equiv & \sum_{a_{i+1},\ldots,a_{n}}f(a_{1},\ldots,a_{i},a_{i+1},\ldots,a_{n},t)
\end{eqnarray*}

Let's mark that under $f^{id}(x)=const$ the equation (\ref{Dyn})
coincides in form with the equation of the random walk on the ultrametric
tree which is the Kolmogorov \& Feller's equation \cite{G} for the
homogenous Markov's proseses distribution function. The last equation
in the limit $n\rightarrow\infty$ becomes the ultrametric equation
of the random walk on the $p$-adic numbers field (Vladimirov's equation)
\cite{ABKO,VVZ}. We have to mark that in our interpretation the function
$f(x)$ is not density of the distribution probability. It determines
a point of the system configurational space which is the space of
functions $f(x)$ satisfying the condition $\sum_{x\in U_{n}}f(x)=1$.
In this case $f(x,t)$ has determined the path in the configurational
space and therefore the equation (\ref{Dyn}) describes the deterministic
dynamics of the fractal-cluster system.

\begin{figure}[ht]
\begin{centering}
\includegraphics[clip,width=0.7\linewidth]{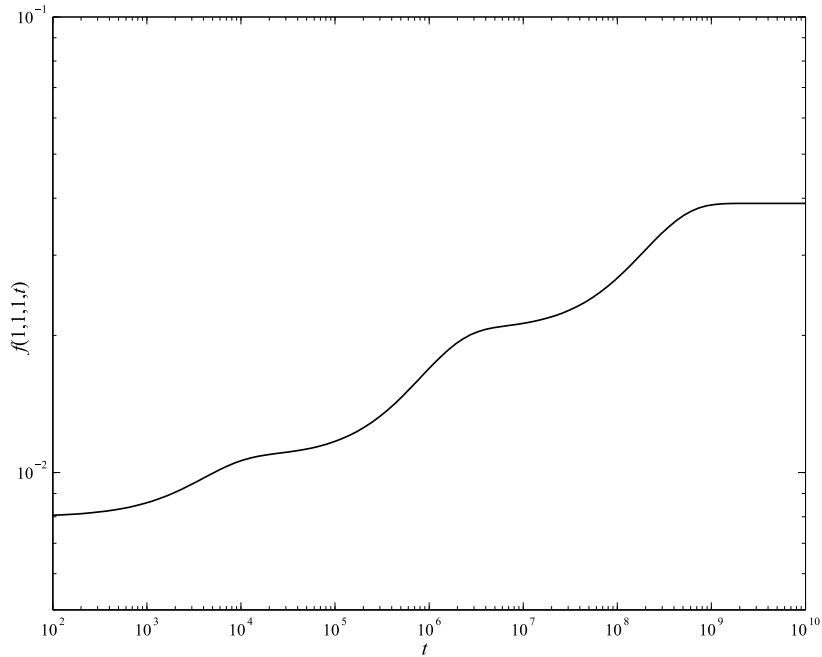} \caption{Dependence $f(1,1,1,t)$ for $p=5,$ $n=3$, $b=4$, $T=1$. Here
are clearly visible characteristic transition times in clusters: $\tau_{3}\sim10^{3}$,
$\tau_{2}\sim10^{6}$, $\tau_{1}\sim10^{9}$. }

\par\end{centering}

\label{fig}
\end{figure}

Let discuss the equation (\ref{Dyn}) solution for the $3$-levels
system ($n=3$) with self-similar (\ref{ss}) and homogenous ($q_{a}=\dfrac{1}{5},\: a=1,\ldots5$)
initial distribution on the clusters. We choose the function $K\left(\lambda\right)$,
which determines the resource transfers from the highest level cluster
$y$ to the any highest level cluster $x$ per the time unite in the
following form:
\begin{equation}
K\left(\lambda\right)=\dfrac{1}{T\lambda^{b}},\label{K}
\end{equation}
where $b$ is the model parameter, which has characterized "an activity"
of the resource redistribution, $T$ is a parameter which determines
the time scale. Let the initial distribution is uniform:
\begin{equation}
f(a_{1},a_{2},\ldots,a_{n},0)=\dfrac{1}{5^{n}}.\label{ic1}
\end{equation}
We will be interested by the resource distributional dynamics in the
first level clusters i.e. the functions
\[
f(a,t)=\sum_{a_{2},a_{3,}\ldots,a_{n}}f(a,a_{2},a_{3},\ldots,a_{n},t).
\]
The solution of the equation (\ref{Dyn}) with the initial condition
(\ref{ic1}) found numerically. On fig.\ref{fig} the resource value
from the time trend $f(1,1,1,t)$ in the subcluster of highest level
$(1,1,1)$ for the $3$-level system under certain fixed values of
the model's parameters has been presented. It is clear that transfer
of this cluster to the ideal state is realized on the character times
by order $\tau_{3}\sim10^{3}$. So far, on the character times by
order $\tau_{2}\sim10^{6}$ are realized transfers to the ideal state
of the second level subclusters and at the end on the character transfer
times by order $\tau_{1}\sim10^{9}$ are realized by transfers to
the ideal state of the first level subclusters, i.e. transfer of the
whole system.

The practical application of this model demands of the its parameter's
identification. In this case, the function $K(\lambda)$ choice depends
from the system type and have to be determined by an empirical way.
For example, besides the power form, it is possible to derive a choice
of the function (\ref{K}) in exponential $\dfrac{1}{T\exp\left(b\lambda\right)}$
or in logariphmic $\dfrac{1}{T\log\left(1+b\lambda\right)}$ forms.
For the adequate choice of this function it is necessary for each
researched system to have empirically determined values of the characteristic
transition times $\tau_{i}$ $i=1,\ldots,n$ for the clusters of all
level to the equilibrium state. For example for the 3--level system
under three known conditions, the empirical values $\tau_{1}$, $\tau_{2}$
and $\tau_{3}$ with the help (\ref{tau}) it is possible to realize
the optimal choice of the function $K(\lambda)$ type and identification
of its two parameters -- $T$ and $b$.

\section{Conclusion}

\label{sect:3}

In this article the scenario of the resource distributional dynamics
in those closed systems which relate to "an organism" class has
been suggested. On the basis of the natural proposals the evolution
equation for the distribution function in such a system has been written.
Solutions of this equation for large time pass into the stationary
solution corresponding to the ideal distribution of resources. The
main principle of dynamics in the closed fractal-cluster system construction
is the fact that the highest levels space of the clusters has the
indexed hierarchical structure which generates the ultrametric structure.
The general suggestion consists of the fact that dynamics of the resource
distribution in the closed systems is determined by the whole ultrametric
structure of the highest levels of the cluster's space. Each cluster
has its` own characteristic time of transition to the ideal state.
This time is determined by the ultrametric size of the cluster (it
is a maximum distance $d(x,y)$ between the higher level clusters
$x$ and $y$, which include to this cluster through this function
$K\left(d(x,y)\right)$, and which determines the value of the resource,
which transfers the resource from the higher level cluster y in any
highest level cluster $x$. We have a wide range of choice of the
function $K\left(d(x,y)\right)$. Under numerical analysis we use
the power form of this function but we can choose the logarithmic
and other forms of the one. In our opinion this choice has to be special
for each particular system. Our suggestion is determined by the mechanics
of the resource transfer dynamics in the closed system: from the beginning,
the highest level clusters transfer to the ideal state then the clusters
of lower do it and so on. Therefore one always has the hierarchy of
characteristic transition times to the ideal state for all clusters.
This hierarchy is determined by the whole system dynamics. This fact
is demonstrated by the numeric solutions the equation for the resource
distribution function. The particular form of the function $K\left(d(x,y)\right)$
depends from the type of the modeling system and it must be determined
on the best approximation of the characteristic transition time to
the equilibrium state for the clusters of the all levels.

\bigskip{}

This work was partially supported by Russian Foundation for Basic
Research grant (project 13-01-00790-a).

\end{document}